\documentclass[11pt]{article}
\usepackage[final]{acl}
\usepackage{times}
\usepackage{latexsym}
\usepackage[T1]{fontenc}
\usepackage[utf8]{inputenc}
\usepackage{microtype}
\usepackage{inconsolata}
\usepackage{graphicx}
\usepackage{pgfplots}
\usepackage{booktabs}
\usepackage{float}
\usepackage{balance}
\usepackage{tikz}
\usetikzlibrary{positioning,arrows.meta}
\pgfplotsset{compat=1.18}

\title{Improving LLM-Assisted Secure Code Generation through\\
Retrieval-Augmented-Generation and Multi-Tool Feedback}

\author{
  Vidyut Sriram \\
  Pennsylvania State University \\
  \texttt{vps5353@psu.edu}
  \And
  Sawan Pandita \\
  Pennsylvania State University \\
  \texttt{sbp5882@psu.edu}
  \And
  Achintya Lakshmanan\\
  Pennsylvania State University \\
  \texttt{aql6062@psu.edu}
  \AND
  Aneesh Shamraj\\
  Pennsylvania State University \\
  \texttt{avs8051@psu.edu}
  \And
  Suman Saha \\
  Pennsylvania State University \\
  \texttt{sumsaha@psu.edu}
}
\begin{document}
\maketitle

\begin{abstract}
Large Language Models (LLMs) can generate code but often introduce security vulnerabilities, logical inconsistencies, and compilation errors. Prior work demonstrates that LLMs benefit substantially from structured feedback, static analysis, retrieval augmentation, and execution-based refinement. We propose a retrieval-augmented, multi-tool repair workflow in which a single code-generating LLM iteratively refines its outputs using compiler diagnostics, CodeQL security scanning, and KLEE symbolic execution. A lightweight embedding model is used for semantic retrieval of previously successful repairs, providing security-focused examples that guide generation. Evaluated on a combined dataset of 3{,}242 programs generated by \textbf{DeepSeek-Coder-1.3B} and \textbf{CodeLlama-7B}, the system demonstrates significant improvements in robustness. For DeepSeek, security vulnerabilities were reduced by 96\%. For the larger CodeLlama model, the critical security defect rate was reduced from 58.55\% to 22.19\%, highlighting the efficacy of tool-assisted self-repair even on ``stubborn'' models.
\end{abstract}

\section{Introduction}

Large Language Models (LLMs) have rapidly advanced the state of automated code generation, yet their outputs remain highly unreliable without structured external guidance. Numerous studies show that LLM-generated programs frequently suffer from compilation failures, logical inconsistencies, and severe security vulnerabilities. For example, \citet{pearce2022copilotsecurity} demonstrate that GitHub Copilot produces insecure code in over 40\% of tested cases, often introducing buffer overflows, unsafe input handling, or cryptographic misuse. These findings indicate that LLMs tend to replicate insecure patterns from training data unless augmented with explicit safety mechanisms.

A broader perspective is offered by \citet{zheng2025survey}, who provide the most comprehensive taxonomy of LLM-based Automated Program Repair (APR) to date. Their survey of sixty-three APR systems concludes that single-pass LLM output is rarely sufficient for correctness or robustness. Instead, the most successful systems incorporate external tools---such as static analyzers, execution engines, and retrieval modules---into procedural repair workflows. This observation motivates our approach: rather than treating the LLM as a standalone generator, we embed it within a multi-step verification and self-repair loop.

Execution-based refinement has also proven effective in improving code reliability. \citet{xu2023selfdebug} introduce Self-Debug, a system where compiler and runtime error messages are fed back into the model to iteratively repair incorrect programs. Their results reveal dramatic improvements in correctness through structured feedback, supporting the hypothesis that LLMs can self-correct when provided actionable diagnostic signals. Complementary work on iterative reasoning, such as STaR by \citet{zelikman2022star}, further demonstrates that multi-step refinement can substantially enhance model performance across domains.

Motivated by these findings, we design a retrieval-augmented, tool-guided self-repair workflow for secure code generation across two contrasting LLMs: DeepSeek-Coder-1.3B and CodeLlama-7B.

Our approach is distinguished by the tight integration of three complementary mechanisms: retrieval-augmented context, multi-tool verification, and iterative self-repair within a single inference-time loop. Retrieval provides prior examples of successful secure repairs, reducing the tendency to repeat unsafe idioms; tool-based analysis supplies concrete diagnostics that expose syntactic, semantic, and security flaws; and iterative self-repair enables the model to adapt its output in response to these signals without requiring fine-tuning or additional generative agents. By combining these signals, the workflow addresses distinct failure modes that are not adequately handled by any component in isolation, offering a practical alternative to agent-heavy or training-intensive approaches.
\section{Literature Review}

The literature consistently emphasizes two key failures of large language models in code generation: (1) the tendency to generate syntactically incorrect or uncompilable code, and (2) the frequent introduction of subtle or severe security vulnerabilities.

\citet{pearce2022copilotsecurity} conducted one of the first systematic evaluations of LLM security, showing that Copilot-generated code commonly contains memory-safety violations, unchecked input flows, integer overflows, and other critical defects. Their vulnerability taxonomy motivated our integration of CodeQL, which detects precisely these classes of issues.

Beyond security defects, several works show that LLMs struggle with deeper program semantics. \citet{xu2023selfdebug} demonstrated that symbolic execution and compiler feedback dramatically improve correctness when used iteratively. Their ``Self-Debug'' findings guided our integration of KLEE, which exposes defects unreachable by simple test-case evaluation.

\citet{zelikman2022star} introduced STaR, a method showing that LLMs substantially improve when placed in iterative refinement loops rather than one-shot inference. This supported our design philosophy that secure code generation requires multi-step reasoning rather than single-pass outputs.

Finally, the comprehensive taxonomy provided by \citet{zheng2025survey} categorizes 63 APR systems and concludes that the most successful ones are tool-augmented rather than purely prompt-driven. They identify retrieval-augmented repair and static-analysis-guided workflows as particularly promising directions. Their findings directly support our integration of RAG with CodeQL and KLEE, highlighting that reliable code generation requires multi-tool orchestration rather than pure LLM sampling.

\section{Methodology}

Our system integrates compiler diagnostics, symbolic execution, static analysis, and retrieval-based guidance into a unified self-repair framework centered around a single code-generating LLM. A lightweight sentence embedding model (\texttt{all-MiniLM-L6-v2}) performs semantic retrieval over a database of previously successful repairs, returning security-focused examples that are injected into the prompt. The retrieval model does not generate or modify code; it serves solely to select relevant contextual examples.

The active LLM (DeepSeek-Coder-1.3B or CodeLlama-7B, depending on the experiment) generates candidate code conditioned on the task description and retrieved examples. The generated program is analyzed by compiler diagnostics (GCC), symbolic execution (KLEE), and static analysis (CodeQL). If any errors are detected---including compilation failures, failing KLEE assertions, or CodeQL security alerts---the diagnostic messages are appended to the prompt and fed back to the same LLM in an iterative self-repair loop. This process repeats for up to three attempts per prompt.

When a program passes all checks, the final code and its associated context are embedded and stored back into the RAG database. During future generations, semantically similar successful repairs are retrieved and used to guide the LLM toward more secure and robust patterns from the outset.

\begin{figure*}[t]
\centering
\includegraphics[width=0.8\textwidth]{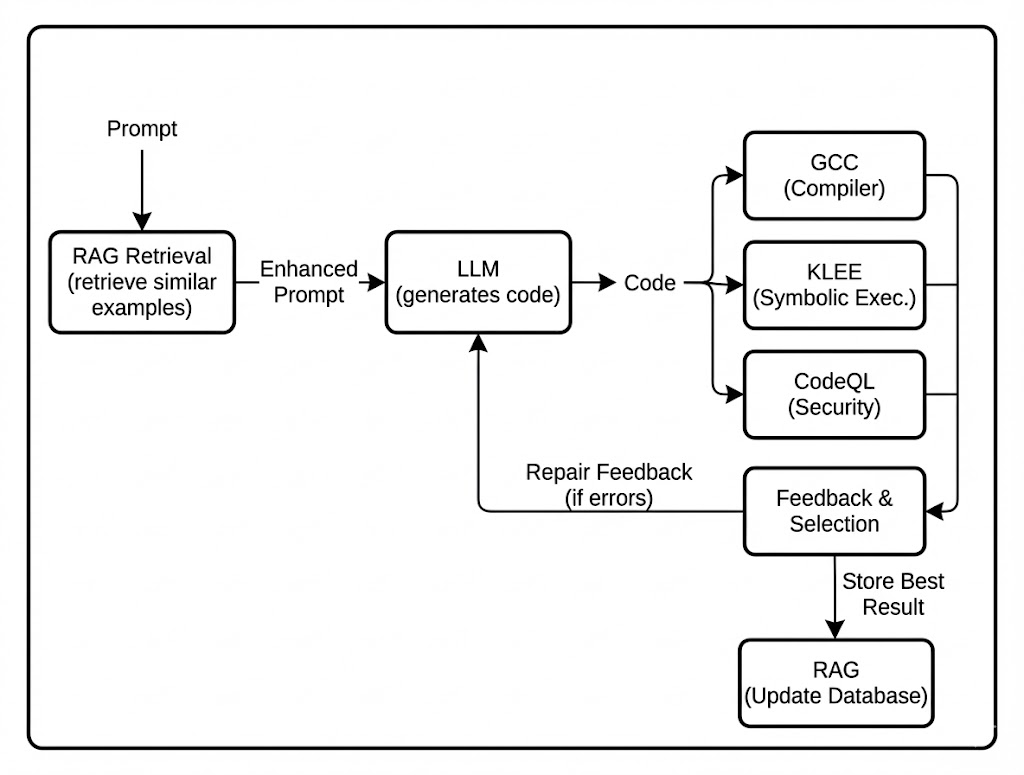}
\caption{Block diagram of the proposed multi-tool feedback system. A lightweight embedding model retrieves security-focused examples from a RAG memory to augment the prompt. A single code-generating LLM produces candidate code, which is analyzed by compiler diagnostics (GCC), symbolic execution (KLEE), and static analysis (CodeQL). Diagnostic feedback is iteratively fed back to the same LLM, and successful repairs are stored back into the RAG database.}
\label{fig:workflow}
\end{figure*}

Figure~\ref{fig:workflow} summarizes this workflow, showing how retrieval, compilation, symbolic execution, and security analysis interact in a closed feedback loop around a single LLM.

\section{Error Taxonomy and Dataset}

We evaluated the workflow on a total of 3{,}242 generated programs (1{,}522 from DeepSeek and 1{,}720 from CodeLlama). Each sample is labeled along three error dimensions:

\begin{itemize}
    \item \textbf{Compilation Errors:} The program fails to compile due to syntax or type issues.
    \item \textbf{Security Errors:} CodeQL reports at least one critical security issue (such as \texttt{cpp/unbounded-write}, unchecked input flows, or integer overflows).
    \item \textbf{Semantic Errors:} The program compiles but fails KLEE-based semantic checks in the provided test harness.
\end{itemize}

\section{Results}
\label{sec:results}

\subsection{Performance on DeepSeek-Coder-1.3B}

With the full workflow enabled for the DeepSeek model (1{,}522 programs), we observed dramatic improvements. The compilation success rate improved by 19.4 percentage points. Most notably, the system achieved a \textbf{98.6\% security-clean rate} (up from a baseline of roughly 64\%), effectively eliminating almost all vulnerabilities.

\begin{table}[t]
\centering
\small
\renewcommand{\arraystretch}{1.2}
\begin{tabular}{lccc}
\toprule
\textbf{Metric} & \textbf{Baseline} & \textbf{Repair Loop} & \textbf{Improvement} \\
\midrule
Compilation & 39.79\% & 20.43\% & 19.36\% Red. \\
Security    & 36.35\% & 1.45\%  & 34.90\% Red. \\
Semantic    & 60.09\% & 5.72\%  & 54.37\% Red. \\
\bottomrule
\end{tabular}
\caption{DeepSeek-1.3B performance comparison.}
\label{tab:deepseek}
\end{table}

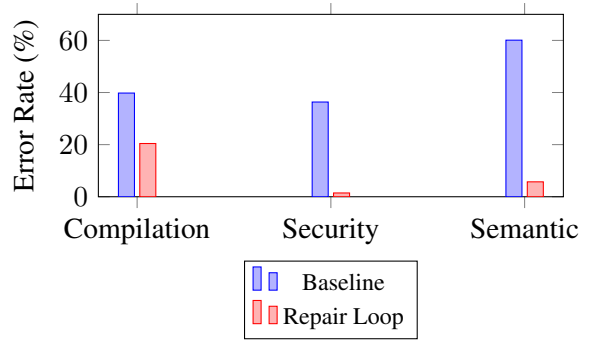
\begin{figure}[t]
\centering
\begin{tikzpicture}
\begin{axis}[
    ybar,
    bar width=6pt,
    ylabel={Error Rate (\%)},
    symbolic x coords={Compilation,Security,Semantic},
    xtick=data,
    ymin=0, ymax=70,
    legend style={font=\footnotesize, at={(0.5,-0.35)},anchor=north},
    width=\columnwidth,
    height=4cm
]
\addplot coordinates {(Compilation,39.79) (Security,36.35) (Semantic,60.09)};
\addplot coordinates {(Compilation,20.43) (Security,1.45) (Semantic,5.72)};
\legend{Baseline,Repair Loop}
\end{axis}
\end{tikzpicture}
\caption{DeepSeek error reduction across categories.}
\label{fig:reduction_deepseek}
\end{figure}

\subsection{Performance on CodeLlama-7B}

We extended the evaluation to the larger CodeLlama-7B model (1{,}720 prompts). As shown in Table~\ref{tab:codellama}, the feedback loop resulted in substantial improvements, though the larger model exhibited distinct behavior compared to DeepSeek.

\begin{table}[t]
\centering
\small
\renewcommand{\arraystretch}{1.2}
\begin{tabular}{lccc}
\toprule
\textbf{Metric} & \textbf{Deliv 1} & \textbf{Deliv 2} & \textbf{Improvement} \\
\midrule
Compilation & 49.71\% & 28.72\% & 20.99\% Red. \\
Security    & 58.55\% & 22.19\% & 36.36\% Red. \\
Semantic    & 19.98\% & 12.18\% & 7.80\% Red. \\
\bottomrule
\end{tabular}
\caption{CodeLlama-7B performance comparison.}
\label{tab:codellama}
\end{table}

\textbf{Analysis:} The RAG system effectively mitigated CodeLlama's tendency to write uncompilable code, cutting the failure rate nearly in half. In terms of security, the baseline CodeLlama model was identified as ``logically correct but dangerously insecure'' (58.55\% defect rate). The feedback loop successfully reduced this to 22.19\%.

\begin{figure}[t]
\centering
\begin{tikzpicture}
\begin{axis}[
    ybar,
    bar width=6pt,
    ylabel={Error Rate (\%)},
    symbolic x coords={Compilation,Security,Semantic},
    xtick=data,
    ymin=0, ymax=70,
    legend style={font=\footnotesize, at={(0.5,-0.35)},anchor=north},
    width=\columnwidth,
    height=4cm
]
\addplot coordinates {(Compilation,49.71) (Security,58.55) (Semantic,19.98)};
\addplot coordinates {(Compilation,28.72) (Security,22.19) (Semantic,12.18)};
\legend{Baseline,Repair Loop}
\end{axis}
\end{tikzpicture}
\caption{CodeLlama-7B error reduction across categories.}
\label{fig:reduction_codellama}
\end{figure}
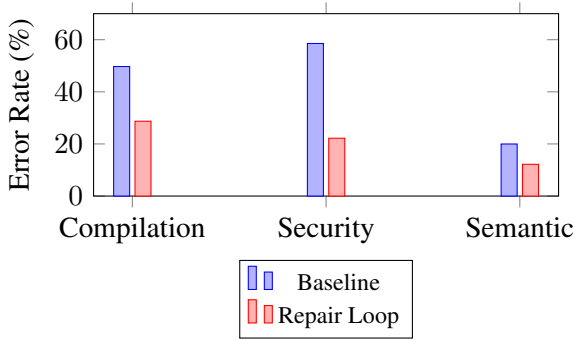

\subsection{Comparative Analysis: The ``Stubborn Model'' Phenomenon}

Comparing the two models reveals a key insight. DeepSeek (1.3B) began with simpler but less secure code, making it highly malleable under tool feedback, achieving near-perfect security (1.45\% error). In contrast, CodeLlama (7B) produced more complex, highly insecure patterns and retained a 22.19\% security error rate even after repair. 

\begin{table}[t]
\centering
\small
\renewcommand{\arraystretch}{1.2}
\begin{tabular}{lcc}
\toprule
\textbf{Final Error Rate} & \textbf{DeepSeek (1.3B)} & \textbf{CodeLlama (7B)} \\
\midrule
Compilation & 20.43\% & 28.72\% \\
Security    & \textbf{1.45\%} & \textbf{22.19\%} \\
Semantic    & 5.72\% & 12.18\% \\
\bottomrule
\end{tabular}
\caption{Final system performance comparison (Deliverable 2).}
\label{tab:final_compare}
\end{table}

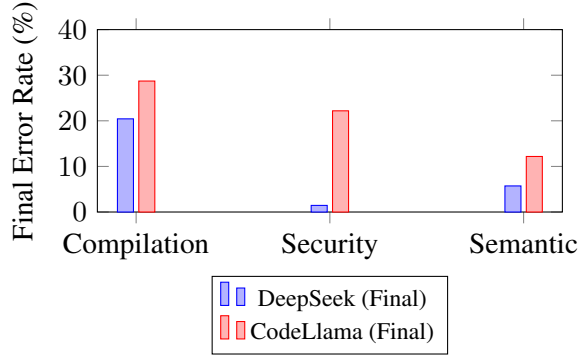
\begin{figure}[t]
\centering
\begin{tikzpicture}
\begin{axis}[
    ybar,
    bar width=6pt,
    ylabel={Final Error Rate (\%)},
    symbolic x coords={Compilation,Security,Semantic},
    xtick=data,
    ymin=0, ymax=40,
    legend style={font=\footnotesize, at={(0.5,-0.35)},anchor=north},
    width=\columnwidth,
    height=4cm
]
\addplot coordinates {(Compilation,20.43) (Security,1.45) (Semantic,5.72)};
\addplot coordinates {(Compilation,28.72) (Security,22.19) (Semantic,12.18)};
\legend{DeepSeek (Final),CodeLlama (Final)}
\end{axis}
\end{tikzpicture}
\caption{Comparative performance of the final system.}
\label{fig:final_compare}
\end{figure}

\section{Limitations and Future Work}

Despite the strong quantitative gains, our study has several limitations. First, the evaluation is restricted to a single problem distribution: short C/C++ programs drawn from competitive-programming style tasks. Real-world software systems exhibit richer structure, larger codebases, and long-lived stateful interactions that may surface different classes of security vulnerabilities. As a result, the reported error reductions may overestimate performance in large industrial codebases.

Second, our workflow focuses on three analysis channels---compilation, CodeQL, and KLEE---whose coverage is inherently incomplete. CodeQL rules target a finite set of vulnerability patterns, and KLEE explores only the paths reachable under a given harness and resource budget. Undetected bugs, side-channel leaks, or concurrency issues remain out of scope. Similarly, we cap the repair loop at three iterations; more steps might further improve robustness at the cost of latency and compute.

Third, our RAG component uses relatively simple retrieval over past repairs. Retrieval is performed using a lightweight embedding model (\texttt{all-MiniLM-L6-v2}), which selects semantically similar successful repairs but does not generate or modify code. We do not yet learn which fixes generalize best, nor do we adapt retrieval based on vulnerability type, program structure, or tool feedback. In addition, we evaluate offline on static benchmarks and do not measure the impact of our pipeline on human developers, such as trust, usability, or debugging time.

\paragraph{Reinforcement Learning for Adaptive Repair Policies.}
A promising future direction is to incorporate reinforcement learning to optimize the repair policy using the tool feedback signals already produced by our pipeline. Compilation success (GCC), semantic correctness (KLEE), and security cleanliness (CodeQL) naturally define sparse but meaningful reward signals, while the number of repair iterations provides a direct latency and compute penalty. Rather than fine-tuning the underlying model weights, reinforcement learning could be applied at the orchestration level to learn adaptive repair strategies, such as which diagnostic signals to prioritize, how aggressively to modify code (e.g., minimal patches versus structural changes), and when to terminate early. This approach would allow the system to balance correctness, security, and efficiency while remaining grounded in objective, tool-based verification.

\paragraph{Agentic Extensions and Multi-Model Scaling.}
Agents or additional models may become valuable when scaling to large, real-world software projects with many files, complex dependencies, and long-horizon interactions. In such settings, agentic components could assist with repository navigation, dependency analysis, tool orchestration, and change planning (e.g., selecting which files to modify and in what order) while keeping edits small and verifiable. However, to preserve safety and prevent unnecessary refactors, any agent-based extension should remain tightly constrained by the same verification tools (build/tests, CodeQL, and symbolic or dynamic checks) and should focus on coordination and planning rather than unconstrained code generation.

Future work will address these limitations by (i) extending the evaluation to multi-file, multi-language codebases; (ii) incorporating additional analyses such as dynamic fuzzing and taint tracking; (iii) learning adaptive repair policies that select tools and prompts based on error type; and (iv) running user studies that place the system inside real development workflows.

\section{Conclusion}

Our results show that retrieval-augmented, multi-tool feedback dramatically enhances LLM-based code generation across model sizes. The workflow reduced DeepSeek's security vulnerabilities by 96\% and CodeLlama's by over 36\%. These findings confirm prior work on APR and LLM debugging: models achieve their best performance when embedded inside tool-governed, feedback-rich pipelines rather than used as standalone generators. Future work will explore larger models, tighter integration between retrieval and static analysis, and reinforcement learning objectives that directly optimize for security and semantic correctness.

\bibliography{custom}

\end{document}